# Resistive switching in reverse: voltage driven formation of a transverse insulating barrier

*Pavel Salev[1,*], Lorenzo Fratino[2], Dayne Sasaki[3], Rani Berkoun[2], Javier del Valle[1,†], Yoav Kalcheim[1], Yayoi Takamura[3], Marcelo Rozenberg[2], Ivan K. Schuller[1]*

[1]Department of Physics and Center for Advanced Nanoscience, University of California San Diego, La Jolla, California 92093, USA

[2]Laboratoire de Physique des Solides, CNRS, Université Paris-Sud, Université Paris-Saclay, 91405 Orsay Cedex, France

[3]Department of Materials Science and Engineering, University of California Davis, Davis, California 95616, USA

*Corresponding author: psalev@ucsd.edu

[†]Present address: Department of Quantum Matter Physics, University of Geneva, 24 Quai Ernest-Ansermet, 1211 Geneva, Switzerland

**Application of an electric stimulus to a material with a metal-insulator transition can trigger a large resistance change. Resistive switching from an insulating into a metallic phase, which typically occurs by the formation of conducting filaments *parallel* to the current flow, has been an active research topic. Here we present the discovery of an opposite, metal-to-insulator switching that proceeds via nucleation and growth of an insulating barrier *perpendicular* to the driving current. The barrier formation leads to an unusual N-type negative differential resistance in the current-voltage characteristics. Electrically inducing a transverse barrier enables a novel approach to voltage-controlled magnetism. By triggering a metal-to-insulator resistive switching in a magnetic material, local on/off control of ferromagnetism can be achieved by a global voltage bias applied to the whole device.**

Materials with unique functional electronic properties can replace large sections of complex circuits, thus greatly improving the scalability and energy efficiency of electronic devices (*1–4*). For instance, using materials in which an insulator-to-metal transition can be triggered electrically makes it possible to mimic the diverse behavior of biological neurons in circuits consisting of just a few components (*5–8*). In contrast, tens of conventional CMOS transistors are required to achieve similar functionalities (*9, 10*). Deep understanding of physical properties and their response to external stimuli becomes critical for designing the applications using such advanced electronic materials. A general picture of the resistive switching process from an insulating into a metallic phase in materials such as in $VO_2$ and $V_2O_3$ (*11–14*), $NbO_2$ (*15, 16*), $(Pr,Ca)MnO_3$ (*17, 18*), etc., is well established. Application of an electric field causes local phase transition, due to Joule heating and/or field-induced carrier doping (*14, 16, 19–23*), which results in the formation of percolating metallic filaments inside the insulating matrix serving as conduits for electric current flow. The filament formation causes strong nonlinearities in the current-voltage (I-V) characteristics such as an S-type negative differential resistance (NDR): a part of the I-V curve has negative dV/dI slope making the overall shape to resemble the letter "S". The opposite type of resistive switching, in which an electrical stimulus drives the material from a metallic into an insulating phase, is a rare phenomenon and an understanding of its microscopic process is still lacking. Several works reported that passing a current can trigger a metal-insulator transition (MIT) in select colossal magnetoresistance manganites (*24–28*), but it remains unknown how this process occurs inside the material. In this work, we show that resistive switching from a metal into an insulator phase proceeds via nucleation and growth of an insulating barrier perpendicular to the current flow, in contrast to metallic filamentary percolation. The barrier formation leads to an unusual N-type NDR nonlinearity in the I-V characteristic. Using theoretical analysis, we present evidence that the transverse barrier formation is a universal property of metal-to-insulator switching, making these findings broadly relevant to a whole class of such resistive switching systems.

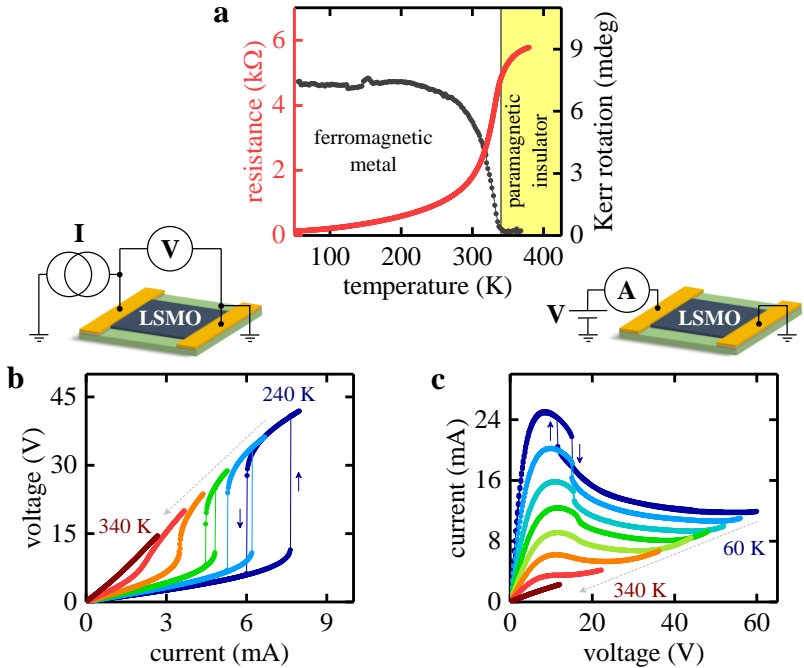

**Fig. 1. Metal-to-insulator resistive switching. a**, Metal-insulator (red line) and magnetic (grey line) transitions in a 50×100 μm² LSMO device probed by electrical transport and MOKE measurements. **b**, Current-controlled I-V curves showing an abrupt and hysteretic metal-to-insulator resistive switching. The measurements are in the 240–340 K temperature range with a step of 20 K (color-coded from blue to dark red). **c**, Voltage-controlled I-V curves showing a gradual resistive switching and an N-type NDR. The measurements are in 60–340 K temperature range with a step of 40 K (color-coded from blue to dark red).

We study metal-to-insulator resistive switching in $La_{0.7}Sr_{0.3}MnO_3$ (LSMO) thin film devices (fabrication details are available in SI 1). Under equilibrium conditions (i.e. without application of high voltage/current), the devices have two coupled phase transitions at $T_c \approx 340$ K: from a low-temperature ferromagnetic metal to a high-temperature paramagnetic insulator (Fig. 1a). The fact that the two transitions go hand-in-hand is the key property that allowed us to map the spatial distribution of resistive switching, as discussed later in the paper.

Resistive switching in LSMO manifests as strong nonlinearities in the current-voltage (I-V) characteristics. Current-controlled I-Vs (Fig. 1b) show an abrupt and hysteretic switching above a certain threshold from a low- to high-resistance state. The switching is volatile, i.e. the device recovers its initial low resistance when the current is ramped down. The switching is observed over a wide temperature range up to $T_c \approx 340$ K where the strong nonlinearities disappear. The high resistance state of ~5.5 kΩ attained after the switching remains the same independent of the measurement temperature and it corresponds to the maximum resistance in the equilibrium resistance-temperature dependence (Fig. 1a). All these observations provide strong evidence that passing electric current triggers the MIT in the LSMO devices. From the analysis of switching parameters (SI 2), the resistive switching is most likely driven by Joule heating. The switching power steadily increases with decreasing temperature, while the switching voltage/electric field has a non-monotonic temperature dependence.

While current-controlled I-V curves showed an abrupt resistive switching, biasing the device with a voltage yielded a qualitatively different behavior (Fig. 1c). Voltage-controlled I-V curves are smooth and display an unusual N-type NDR (the I-V shape resemble the letter "N"). The NDR develops when the device actively undergoes resistive switching: as the applied voltage increases, the current becomes smaller, indicating that the resistance progressively becomes larger. Because N-type NDR is very unusual in resistive switching systems (the switching typically leads to an S-type NDR), we have investigated extensively the origin of the NDR in the LSMO devices.

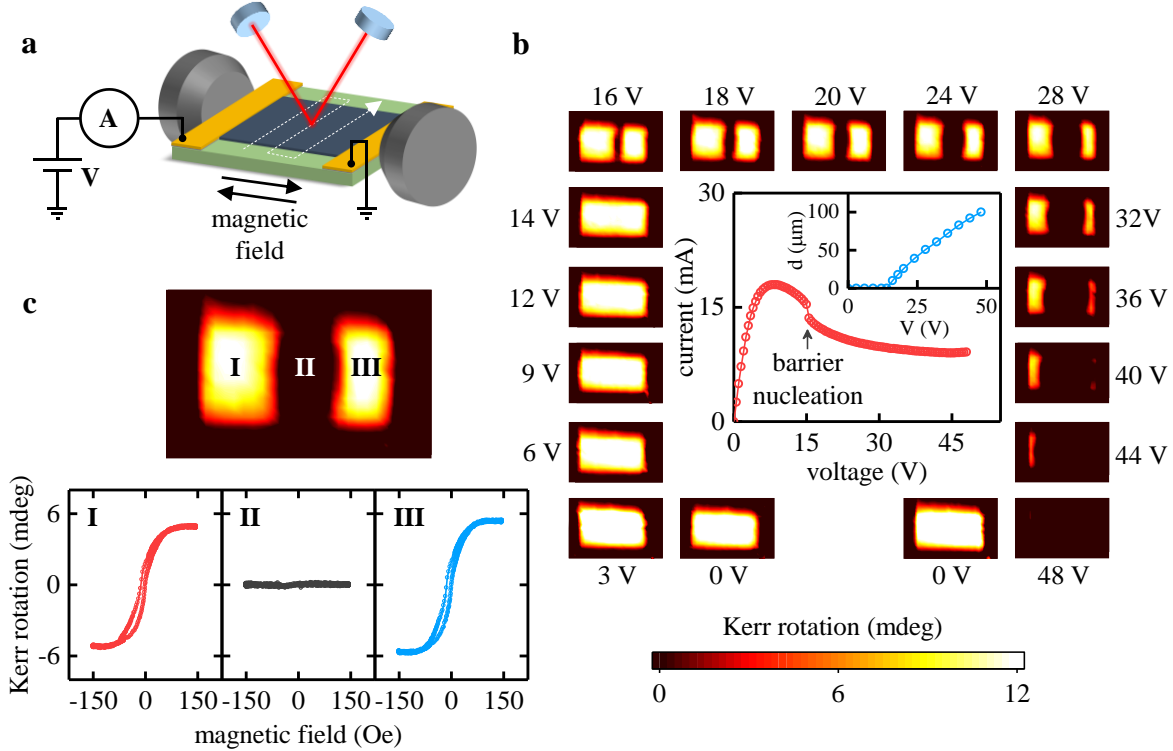

**Fig. 2. Spatial mapping of the metal-to-insulator resistive switching. a**, Schematic of the MOKE measurement setup. The MOKE hysteresis loops were acquired at every *xy*-spot in the device area. The magnetic field was applied in-plane along the device length. Voltage biasing was maintained without interruptions over the entire measurement time. **b**, Simultaneously recorded I-V curve (center) and MOKE *xy*-maps (sides). The bright areas in the maps correspond to the ferromagnetic LSMO. The total field of view is 90×140 µm². In the maps, the electric current flows horizontally. As the I-V progresses through the NDR, a transverse insulating paramagnetic barrier nucleates and grows from the device center. The inset in the I-V plot shows the barrier size, *d*, as a function of applied voltage. **c**, MOKE map and local hysteresis loops corresponding to three device regions (labeled using Roman numerals) recorded at 24 V. While the device sides (regions I and III) show ferromagnetic response, the MOKE signal is zero in the center (region II). All measurements were done at 100 K.

To understand the underlying microscopic mechanism of metal-to-insulator resistive switching, we performed *in-operando* imaging of the LSMO devices exploiting the fact that the MIT occurs simultaneously with the magnetic transition. Using scanning magneto-optical Kerr effect (MOKE) microscopy (Fig. 2a), we mapped the spatial distribution of ferromagnetic regions while applying a voltage bias. The measurement procedure involved recording MOKE hysteresis loops at every spot over the device area using a 5-µm-size laser beam. We represent the data by plotting *xy*-maps of the MOKE loops magnitudes (i.e. Kerr rotation angle), which provides a graphic visualization of the spatial distribution of ferromagnetism. We note that our MOKE maps are different from conventional MOKE images in which the contrast originates from domains of different magnetization orientation. In our case, the bright areas correspond to ferromagnetic regions, while the dark areas indicate the absence of ferromagnetism.

We found that the metal-to-insulator resistive switching occurs via nucleation and growth of an insulating barrier that spans through the entire device width in the direction perpendicular to the electric current flow. Fig. 2b shows the MOKE maps at different voltages and the corresponding I-V curve. The device remains uniformly ferromagnetic (metallic) below 15 V, but applying a higher voltage causes the LSMO to transform into a qualitatively new state. At 16 V, the I-V curve displays a small jump. At the same time, a ~5-µm-wide nonmagnetic domain nucleates in the device center. The domain spans laterally across the full device width and its size, $d$, grows with the applied voltage until it encompasses the entire device at 48 V (inset in the I-V plot in Fig 2b). Because of the coupling between magnetic and transport properties, our measurements imply that the resistive switching from a metal into an insulator does not occur uniformly throughout the device. Instead, an unusual out-of-equilibrium phase separation is favored: a transverse insulating barrier cuts through the conducting matrix and blocks the electric current flow. This is a "mirror image" of the resistive switching from an insulator into a metal (for example, in vanadium oxides (*11–13*)), which proceeds via nucleation and growth of longitudinal metallic filaments that serve as conduits for the current flow. The voltage-induced barrier formation is highly reproducible: multiple devices patterned on multiple LSMO films showed exactly the same behavior (SI 3). On the other hand, we did not find any indications of a barrier appearing during equilibrium thermal transition (without applied voltage), which indicates that the barrier is not simply due to inhomogeneities in the LSMO film (SI 4). Therefore, the nucleation of an insulating barrier is not an accidental device property but rather a general feature of voltage-driven MIT.

Resistive switching in LSMO enables a novel approach to voltage-controlled magnetism. Voltage-driven metal/insulator/metal phase separation during the switching results in an unusual ferromagnetic/paramagnetic/ferromagnetic domain configuration. Fig. 2c shows a MOKE map and three local hysteresis loops recorded in the LSMO device under the application of 24 V. While both left and right device sides (labeled I and III) have normal ferromagnetic loops, the center domain (labeled II) does not show any magnetic response: the recorded MOKE loop is just a flat line. Therefore, the device can serve as an on/off magnetism switch where a local switching is achieved using a global voltage stimulus. For the purpose of imaging, the devices had relatively large dimensions (50×100 µm$^2$) and the minimum observed paramagnetic domain size was ~5 µm (which could be a convolution of the laser beam size). Our analytical analysis suggests that the minimum domain size could be reduced substantially by reducing the device dimensions and by selecting a material with a high insulator to metal resistivity ratio (SI 5). This presents a unique opportunity to achieve an efficient voltage-controlled magnetism at the nanoscale. As discussed later in the paper, we expect the development of a large thermal gradient associated with the barrier formation. Thus, resistive switching in LSMO could enable a novel platform for spin caloritronics (*29*). Because of the fundamental nature of the metal-to-insulator switching, the temperature gradient emerges naturally in simple, planar-geometry LSMO devices biased with a voltage, which eliminates the need for external heaters. This simplifies the control over the temperature profile and at the same time provides an easy access for a variety of surface-sensitive techniques such as magnetic force microscopy, magnetic photoelectron emission microscopy, nitrogen-vacancy centers in diamond, etc.

The insulating barrier formation and the appearance of N-type NDR during the voltage-controlled switching are strongly linked to each other. Neither barrier nor NDR can be observed when the LSMO device is switched using a current bias (Fig. 3). Under current-controlled conditions, the I-V displays an abrupt jump and, at the same time, the magnetic signal in the MOKE maps completely vanishes throughout the entire device. The different behaviors in current- and voltage-controlled switching (full device transition vs. barrier formation) can be understood by considering the impact of the barrier nucleation on the current and voltage distributions within the device. The transverse barrier does not disrupt a homogeneous current distribution because it spans across the full device width. Therefore, the entire device switches at a threshold current under current-controlled conditions. The situation is radically different under voltage-controlled conditions,. The nucleation of an insulating barrier concentrates the applied voltage, which leads to a highly inhomogeneous voltage distribution inside the device. The development of this voltage inhomogeneity explains why the current- and voltage-controlled I-V curves are different (abrupt switching vs. N-type NDR) when the device actively undergoes the resistive switching, but the two curves perfectly coincide when the device is either in a uniform metallic or insulating state (I-V plot in Fig. 3).

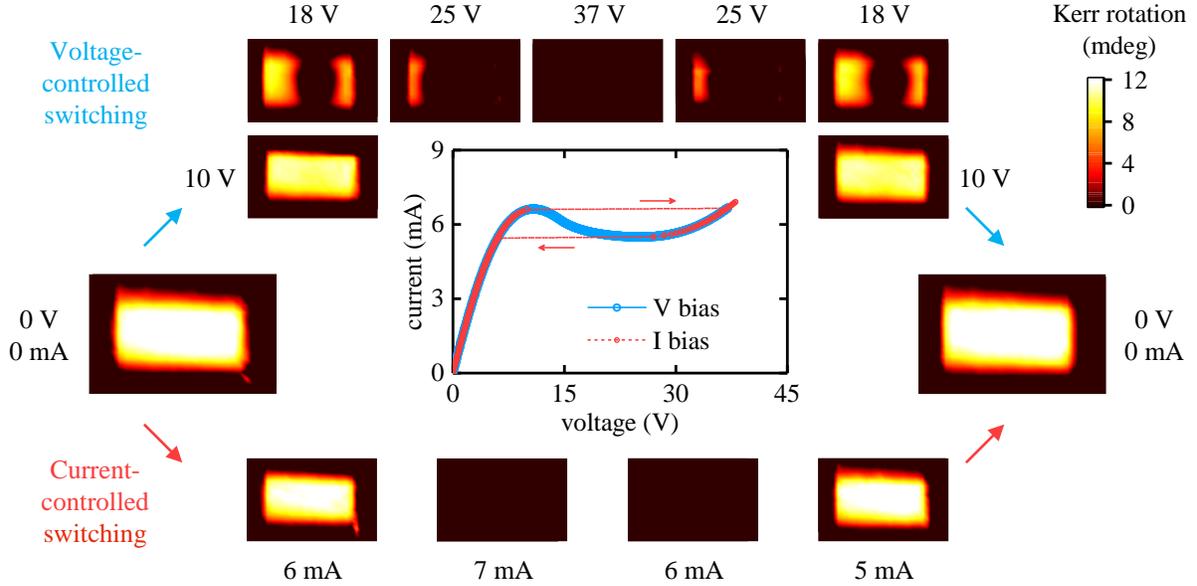

**Fig. 3. Current- vs. voltage-controlled metal-to-insulator switching.** The plot in the center shows two overlaid I-V curves recorded under current- (red) and voltage-controlled (blue) conditions. The I-V curves perfectly coincide outside the NDR region. Corresponding MOKE maps (surrounding the I-V plot) show the switching via an insulating barrier nucleation and growth in the voltage-controlled regime (top maps) and the switching of the entire device at once in the current-controlled regime (bottom maps). In the maps, the current flows horizontally. The field of view is 90×140 µm$^2$. The measurements were done at 250 K.

We achieved an excellent agreement between the experimental results and resistive network simulations (Fig. 4 a and b) by considering a realistic resistance vs. temperature dependence and thermal effects due to Joule heating. The details of simulations are available in SI 6. Fig. 4c shows local temperatures vs. voltage plots at different positions within the device. We found that the applied voltage initially heats up the entire device. This heating, however, is not homogeneous. Just prior to the formation of the insulating barrier, the temperature at the device center is several Kelvins higher compared to the edges (Fig. 4d). This is due to the thermal coupling of the edges to the device electrodes that are at the substrate temperature. Even a small temperature deviation at the center is enough to initiate locally a resistance-power positive feedback loop because of the proximity to the MIT. Higher local temperature increases the resistance, which leads to an increase of the local voltage drop and the power dissipation, further increasing the local temperature. As a result, the temperature at the device center abruptly increases well above the $T_c$ and an insulating barrier nucleates. The barrier concentrates most of the voltage drop, which keeps its temperature high, while the rest of the device cools down almost to the substrate temperature. A distinct and a rather sharp boundary between the two phases can be always observed: the temperature of the metallic regions remains close to the substrate temperature up until these regions are "swallowed" by the growing barrier as the applied voltage increases. The only necessary ingredient enabling the barrier formation is a thermal transition from a low- into a high-resistance state. Because we did not have to make any explicit assumption about phase separation, magnetic properties, defect density profiles, etc., we conclude that our analysis provides a universal description of voltage-triggered metal-to-insulator phase transition mediated by Joule heating. Therefore, many other materials in the manganite family and some magnetic semiconductors could have similar resistive switching behavior.

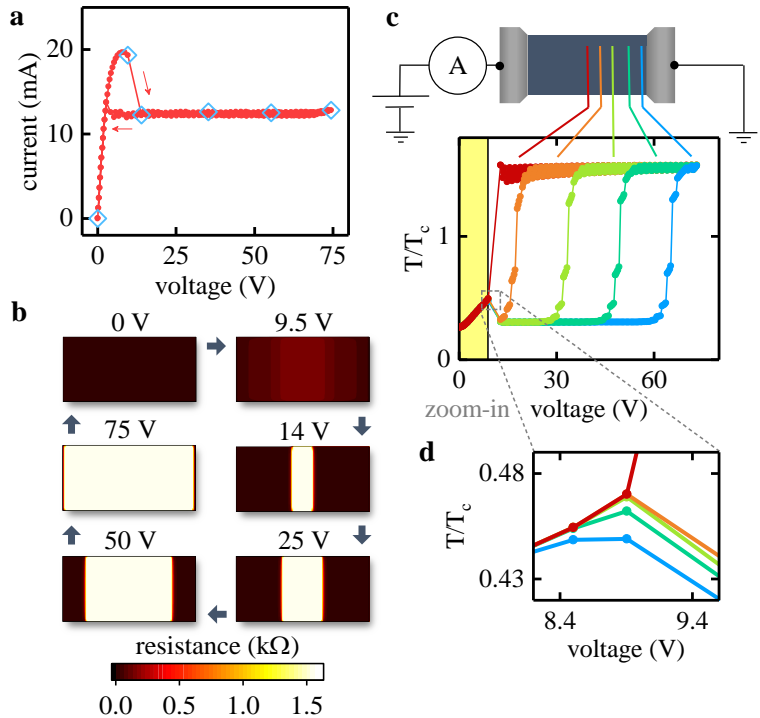

**Fig. 4. Computational analysis of metal-to-insulator resistive switching. a,b** Calculated voltage-controlled I-V curve (**a**) and resistance *xy*-maps (**b**) showing a remarkable agreement with the experimental data. Diamond symbols in **a** highlight the I-V points for which the resistance maps are shown in **b**. **c,** Temperature vs. applied voltage at several positions along the device length color-coded from red at the center to blue near the electrode. The yellow-shaded region corresponds to the state before the barrier formation. When the barrier nucleates, the temperature at the device center abruptly increases above $T_c$, while the rest of the device cools down. As the barrier grows and reaches each point, the temperature at that point increases rapidly above $T_c$. **d,** A zoom-in of the temperature vs. voltage plot at the barrier nucleation. At this magnification, it becomes apparent that just prior to the barrier nucleation, the temperature at the center is higher compared to the edges. This thermal inhomogeneity triggers locally the power-temperature positive feedback loop culminating in the insulating barrier nucleation.

The results presented in this paper complete the picture of resistive switching in MIT materials. Two switching types are possible: insulator-to-metal (I→M), e.g. in $VO_2$ or $NbO_2$, and metal-to-insulator (M→I), e.g. in LSMO as shown in this work. These two switching types have contrasting behaviors. Under voltage biasing, I→M is abrupt and hysteretic while M→I is gradual and has an N-type NDR. Under current biasing, I→M is gradual having an S-type NDR and M→I is abrupt and hysteretic. In both cases, the gradual switching is accompanied by a spatially inhomogeneous transition. In I→M, this inhomogeneity is a conducting filament parallel to the current flow, while in M→I an insulating barrier perpendicular to the current flow forms during the switching. Combining the two types of resistive switching provides a broad range of nonlinear electrical properties, which greatly enriches the design space for complex-behavior electronic devices. In LSMO, where the transport and magnetic properties are coupled, the transverse insulating barrier formation allows local switching of magnetism by an applied voltage. The thermal behavior associated with the barrier has direct relevance to devices based on spin caloritronics effects.

### Acknowledgement

This work was supported as part of the Quantum Materials for Energy Efficient Neuromorphic Computing (Q-MEEN-C), an Energy Frontier Research Center funded by the U.S. Department of Energy, Office of Science, Basic Energy Sciences under Award # DE-SC0019273. MR was partially supported by the ANR project MoMA.

# Resistive switching in reverse: voltage driven formation of an insulating barrier

*Pavel Salev[1], Lorenzo Fratino[2], Dayne Sasaki[3], Rani Berkoun[2], Javier del Valle[1,4], Yoav Kalcheim[1], Yayoi Takamura[3], Marcelo Rozenberg[2], Ivan K. Schuller[1]*

[1]Department of Physics and Center for Advanced Nanoscience, University of California San Diego, La Jolla, California 92093, USA

[2]Laboratoire de Physique des Solides, CNRS, Université Paris-Sud, Université Paris-Saclay, 91405 Orsay Cedex, France

[3]Department of Materials and Science Engineering, University of California Davis, Davis, California 95616, USA

[4]Present address: Department of Quantum Matter Physics, University of Geneva, 24 Quai Ernest-Ansermet, 1211 Geneva, Switzerland

## Supporting Information 1

$La_{0.7}Sr_{0.3}MnO_3$ (LSMO) films of 20 nm and 50 nm thickness were grown epitaxially on a (001)-oriented $SrTiO_3$ (STO) substrates using pulsed laser deposition with a laser fluence of 0.7 J·cm$^{-2}$ and frequency of 1 Hz. During the growth, the substrate temperature was held at 700 °C with an oxygen pressure of 0.3 Torr. After deposition, the films were slowly cooled to room temperature in 300 Torr $O_2$ to ensure proper oxygen stoichiometry.

The structural quality of the synthesized LSMO films was examined using x-ray diffraction. Fig. S1a shows a specular θ-2θ scan in the vicinity of the STO (002) Bragg peak. The LSMO $(002)_{pc}$ peak is at $Q = 3.25$ Å$^{-1}$ giving an out-of-plane lattice constant of 3.86 Å. The film's diffraction pattern has clear Laue oscillations, which attest to a coherent crystal structure with smooth interfaces. Fig. S1b shows a reciprocal space map in the vicinity of the STO (103) peak. The LSMO film peak has the same $Q_x$ component as the STO substrate, indicating that the film is fully strained. Fig. S1c shows the resistance-temperature dependence of the film before patterning (grey line) and of the fabricated devices (green and red). The three curves display the same metal-insulator transition (MIT) in terms of transition temperature and magnitude of the resistance change. This behavior implies that the device fabrication preserved the film's chemical and structural integrity.

The electrodes of (100 nm Au)/(20 nm Pd) for electrical measurements were made using standard photolithography process and e-beam evaporation. The bottom Pd layer was used to achieve a low contact resistance with the LSMO film. After the electrode fabrication, 50×100 μm$^2$ and 50×50 μm$^2$ devices were defined using reactive ion etching in an Ar/Cl$_2$ atmosphere. Fig. S1c shows the resistance-temperature dependence of the film before patterning (grey line) and of the fabricated devices (green and red). The three curves display the same metal-insulator transition (MIT) in terms of transition temperature and magnitude of the resistance change. This behavior implies that the device fabrication preserved the film's chemical and structural integrity.

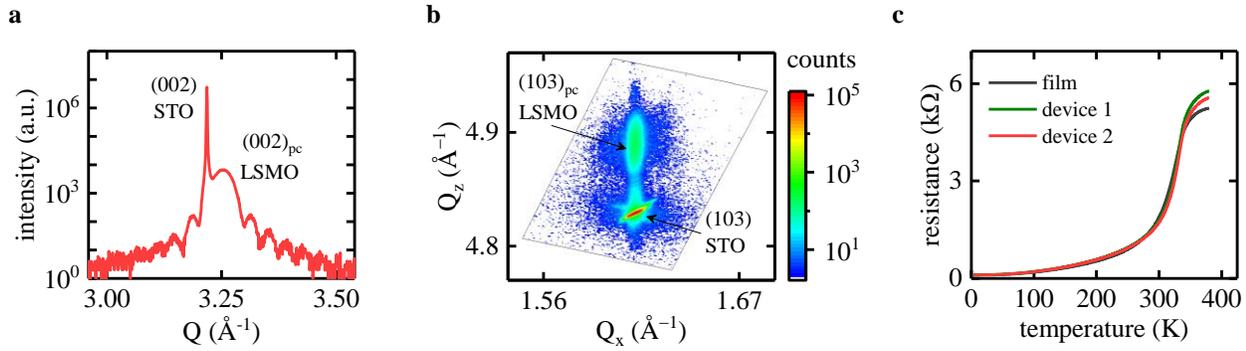

**Fig. S1. a, b,** X-ray diffraction of the LSMO film: specular θ-2θ scan in the vicinity of the STO (002) peak (**a**) and reciprocal space map in the vicinity of the STO (103) peak (**b**). **c,** Resistance vs. temperature curves of the LSMO film (grey line) and two 50×100 μm² devices (green and red lines) showing similar behavior.

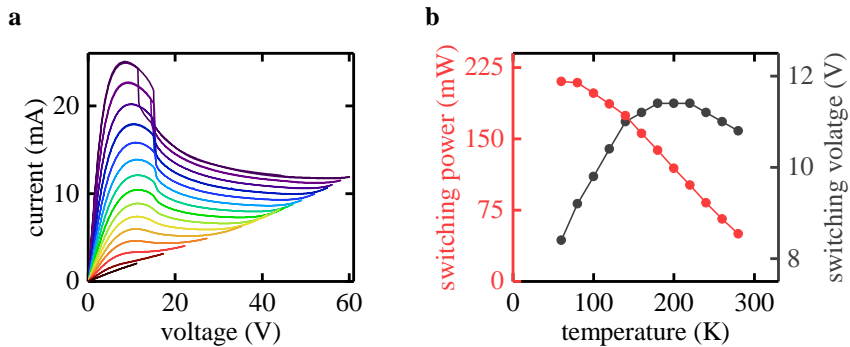

**Fig. S2. a,** Voltage-controlled I-V curves recorded in a 60–340 K range using a step of 20 K. **b,** Switching power and voltage corresponding to the onset of the NDR, extracted from the I-V curves in **a**.

## Supporting Information 2

Resistive switching measurements were performed in a Quantum Design PPMS DynaCool cryostat using a Keithley 2450 source meter in either voltage- or current-controlled mode. Fig. S2a shows voltage-controlled I-V curves recorded in a 60–340 K temperature range. Several I-V curves from this figure are also shown in the main text in Fig. 1c. To characterize the switching parameters, we extracted the currents and voltages corresponding to the onset of negative differential resistance (NDR), i.e. the point at which dV/dI changes sign. Fig. S2b shows the switching power and the switching voltage dependence on temperature. While switching power steadily decreases with increasing temperature, the switching voltage shows a non-monotonic behavior. This result suggests that Joule heating rather than electric field drives the metal-to-insulator switching in the LSMO devices.

## Supporting Information 3

Magneto-optical imaging of resistive switching was performed in a Montana Instruments NanoMOKE 3 system. The light source was a 660 nm laser focused to a 5-μm-size spot. The magnetic field was cycled in the ±150 Oe range at a 4.7 Hz repetition rate. Voltage/current was applied to the LSMO devices using a Keithley 2450 source meter. The measurement procedure was the following. First, a voltage/current was set. Then the laser was focused at a starting *xy*-coordinate and a MOKE hysteresis loop averaged over 20 cycles was recorded. The loop recording was continued at every *xy*-coordinate until the entire imaging area was covered (typically 90×140 μm²). After this, a new voltage/current was set and the loop recording procedure was repeated. We note that the voltage/current was maintained without interruptions during the entire imaging procedure.

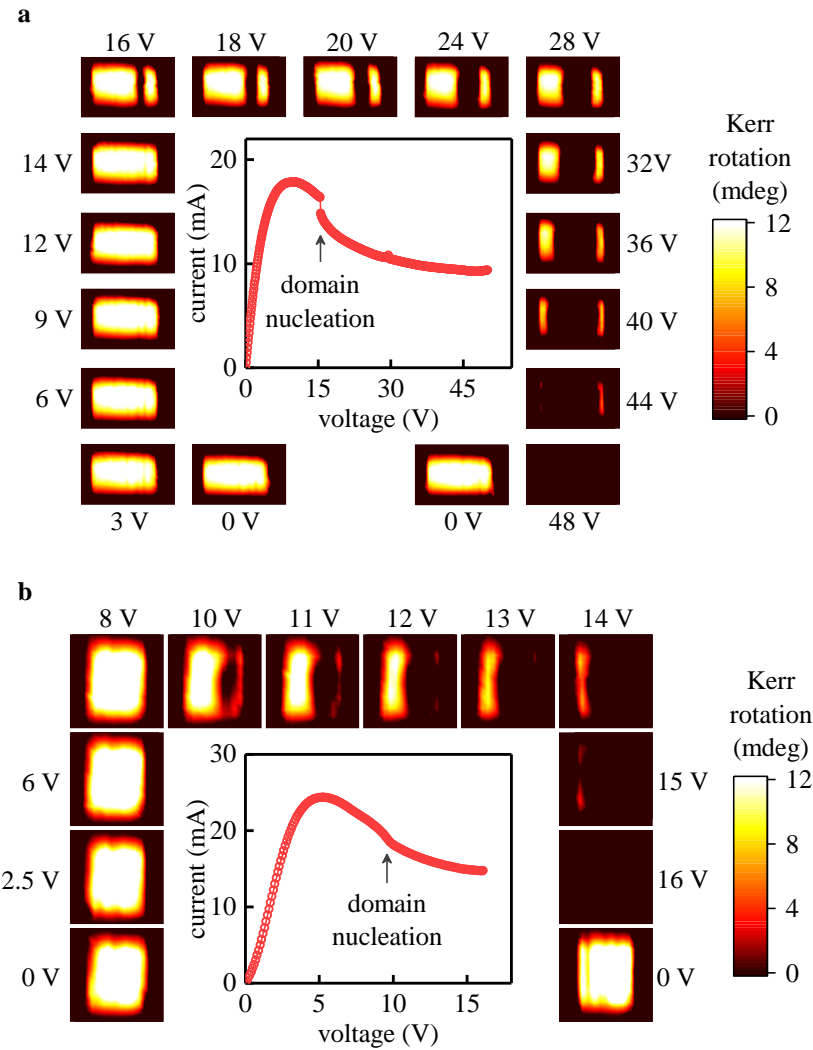

**Fig. S3. a, b**, Simultaneously recorded I-V curve (center) and MOKE amplitude *xy*-maps (sides) in two different samples having different device geometry. Device in **a** is patterned in a 20 nm thick film and has 50×100 μm² size. Device in **b** is patterned in a 50 nm thick film and has 50×50 μm² size. The field of view in the MOKE maps is 90×140 μm² in **a** and 75×75 μm² in **b**. In the maps, the current flows horizontally. the All measurements were performed at 100 K.

Fig. S3 shows the MOKE maps at different voltages and the corresponding I-V curves recorded in the same sample as in the main text but in a different device (panel a) and in another LSMO sample of different film thickness (50 nm) and different device dimensions (50×50 μm², panel b). In both cases, we observed the same behavior as described in the main text: the switching from a metal into an insulator proceeds via nucleation and growth of an insulating barrier that spans through the entire device width in the direction perpendicular to the current flow. The repeatability of the switching behavior indicates that the formation of an insulating barrier is a general property of metal-to-insulator switching.

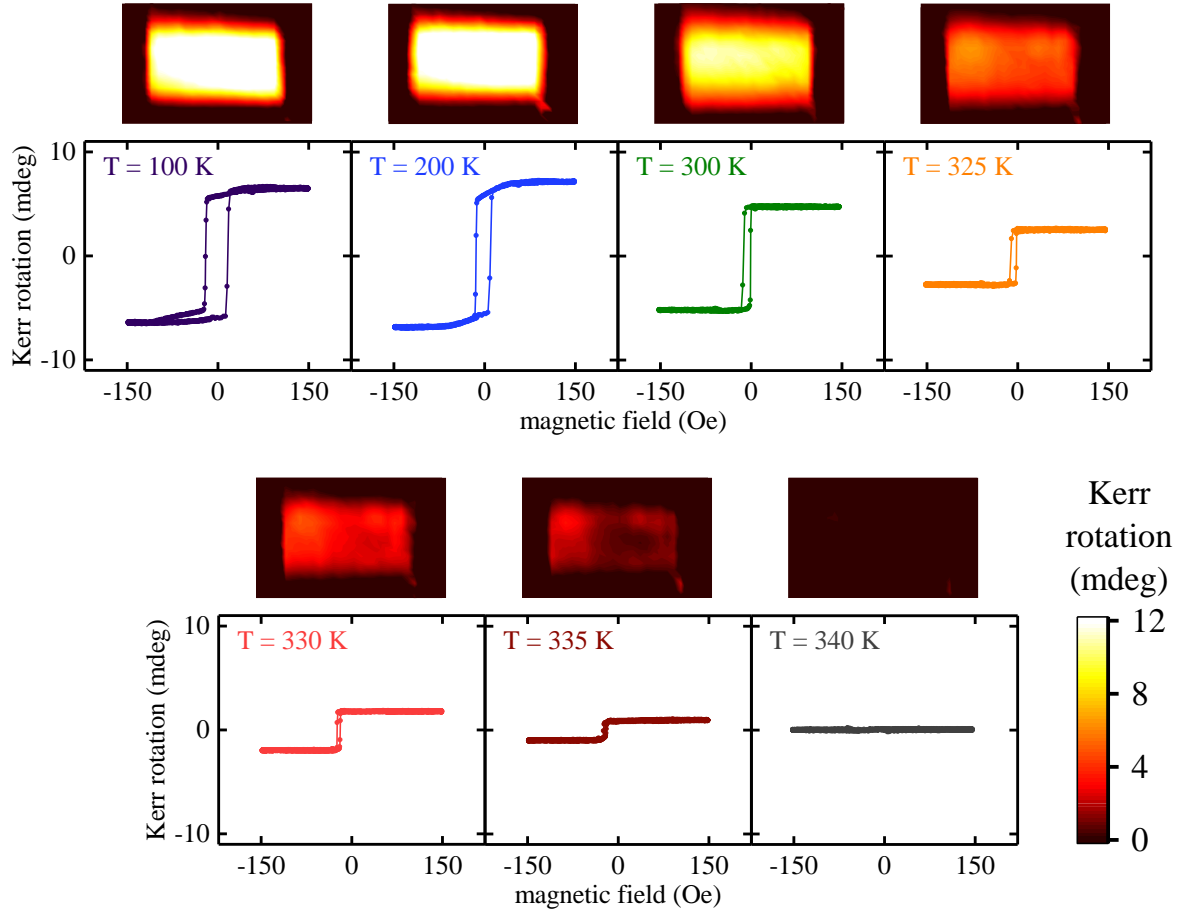

**Fig. S4.** MOKE maps and corresponding MOKE hysteresis loops at different temperatures recorded during the equilibrium thermal transition (i.e. without applying voltage/current to the LSMO device). In contrast with the formation of an insulating barrier, the thermal transition is spatially uniform. The field of view in the MOKE maps is 90×140 µm$^2$.

## Supporting Information 4

To check whether the formation of an insulating barrier is not just an anomaly of the LSMO film, which, for example, could be introduced during the device fabrication, we measured the MOKE maps over 100–400 K temperature range without applying voltage (Fig. S4). We used the same imaging procedure and the same magnetic field settings as described in Supporting Information 3. Under equilibrium conditions, we observed a spatially uniform transition throughout the device. Importantly, we found no correlation to the formation of an insulating barrier that we observed during resistive switching. This result demonstrate that the nucleation and growth of an insulating barrier is a special property of the electrically-driven metal-to-insulator transition.

## Supporting Information 5

To derive an analytical equation for the insulating barrier size we consider a simple model consisting of series resistors that represent metal/insulator/metal configuration and we take into account the heat exchange between the metal regions, insulator region, and the substrate (Fig. S5.1). At equilibrium, i.e. $\partial T/\partial t = 0$, the heat equation for the insulating barrier can be written as

$$\frac{V_{apl}^2 R_I}{(R_M + R_I)^2} = k_s(T_I - T_0) + 2k_f(T_I - T_M) \qquad (5.1)$$

where $R_I$ and $R_M$ are resistances of insulator and metal, $T_I$, $T_M$, and $T_0$ are temperatures of insulator, metal, and substrate, $k_f$ and $k_s$ are thermal conductivities of heat exchange within the film and between the film and the substrate. Introducing material properties and device dimensions as (defined in Fig. S5.1 a)

$$\begin{aligned} R_I &= \rho_I \frac{d}{t \cdot w} \\ R_M &= \rho_M \frac{L - d}{t \cdot w} \\ k_s &= \kappa_s (d \cdot w) \\ k_f &= \kappa_f (t \cdot w) \end{aligned} \qquad (5.2)$$

equation (5.1) can be rewritten in a compact form

$$\boldsymbol{A}\left(\frac{d}{L}\right)^3 + \boldsymbol{B}\left(\frac{d}{L}\right)^2 + \boldsymbol{C}\left(\frac{d}{L}\right) + \boldsymbol{D} = 0 \qquad (5.3)$$

where

$$\begin{aligned} \boldsymbol{A} &= \kappa_s \alpha^2 \\ \boldsymbol{B} &= 2\kappa_s \alpha + 2\kappa_f \alpha^2 \frac{t}{L} \frac{(T_I - T_M)}{(T_I - T_0)} \\ \boldsymbol{C} &= \kappa_s + 4\kappa_f \alpha \frac{t}{L} \frac{(T_I - T_M)}{(T_I - T_0)} - \frac{(\alpha + 1)t}{\rho_M (T_I - T_0) L^2} V_{apl}^2 \\ \boldsymbol{D} &= 2\kappa_f \frac{t}{L} \frac{(T_I - T_M)}{(T_I - T_0)} \\ \alpha &\equiv \frac{\rho_I - \rho_M}{\rho_M} \end{aligned} \qquad (5.4)$$

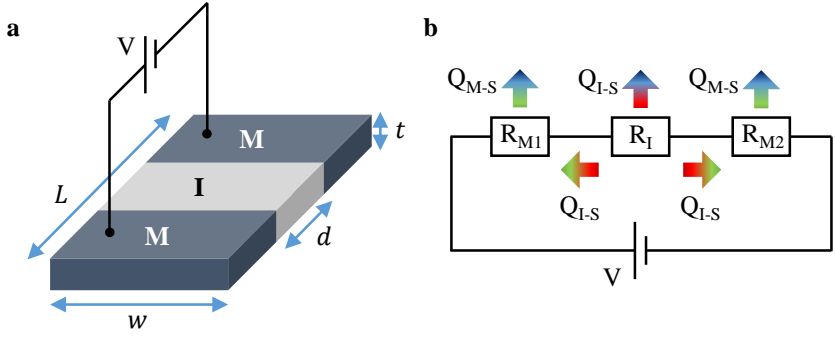

**Fig. S5.1. a,** Device geometry considered in the analytical model: an insulating barrier in the middle separates the two metallic parts. Device dimensions used in the equations are shown. **b,** Schematic of the analytical model showing the electrical circuit and heat exchange flows.

As can be seen from (5.4), all coefficients in (5.3) are positive when $V_{apl}$ is small, therefore there is no positive solution for barrier size, $d$, i.e. the insulating barrier cannot form without the application of a strong enough voltage. We define the $V_{th}$ as the threshold voltage at which (5.3) has the smallest positive solution $d_c$ for the insulating barrier size. Fig. S5.2a shows that at $V_{th}$ the equation (5.3) has a local minimum corresponding to $d_c$. This means that the derivative of (5.3) is also zero at $d_c$, which leads to a system of two equations

$$\begin{cases} \dfrac{d_c}{L} = \dfrac{-C + \sqrt{C^2 - 4BD}}{2B} \\ \dfrac{d_c}{L} = \dfrac{-2B + \sqrt{4B^2 - 12AC}}{6A} \end{cases} \tag{5.5}$$

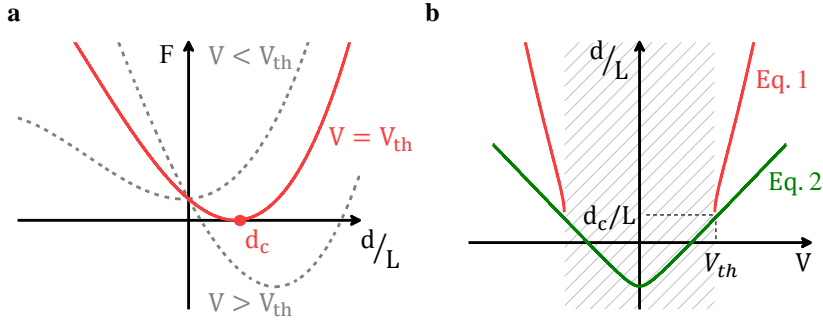

**Fig. S5.2. a,** Schematic plot of equation (5.3) for three voltages: below and above the threshold (grey dashed lines) and at the threshold (red line). The minimum size of the insulating barrier $d_c$ is highlighted. **b,** Schematic plot of system of equations (5.5). The intercept of the two curves corresponds to the minimum insulating barrier size $d_c$ and threshold voltage $V_{th}$ to induce such a barrier. The shaded area highlights the region where Eq. 1 becomes imaginary.

The above equations were derived assuming $d_c \ll L$ in order to obtain analytically solvable equations by dropping $(d_c/L)^3$ terms. Because of the missing cubic terms, system (5.5) actually does not have an exact solution. Fig. S5.2b shows graphically that close to the intercept point, the first equation in (5.5) does not have real values because $\sqrt{C^2 - 4BD}$ becomes imaginary. However, the two curves in Fig. S5.2b come very close to the interception point. Therefore, we can write a condition for the approximate solution as following

$$\begin{cases} \dfrac{d_c}{L} = -\dfrac{C}{2B} \\ C^2 - 4BD = 0 \end{cases} \tag{5.6}$$

System of equations (5.6) contains four unknowns: $d_c$, $V_{th}$, $T_I$, and $T_M$. In order to obtain an approximate solution, we used $T_c$ and $T_0$ as estimations for the insulator and metal regions temperatures, $T_I$ and $T_M$. As we found in our numerical simulations (see Fig. 3 in the main text), such estimations are justified. In addition, the temperature coefficients in (5.4), $(T_I - T_0)$ and $(T_I - T_M)$, are always multiplied by other parameters, such as $\alpha$, $\kappa_f$, $\rho_M$. It is possible to "absorb" the modest deviations of $(T_I - T_0)$ and $(T_I - T_M)$ from $(T_c - T_0)$ into other parameters. Therefore, using $T_I \approx T_c$ and $T_M \approx T_0$ does fundamentally alter the physics of the problem, but greatly simplifies the analytical solution. System of equations (5.6) leads to the following expressions for the insulating barrier size $d_c$ and the threshold voltage $V_{th}$

$$d_c \approx L \sqrt{\dfrac{1}{\alpha\left(\alpha + {k_s L}/{k_f t}\right)}} \tag{5.7}$$

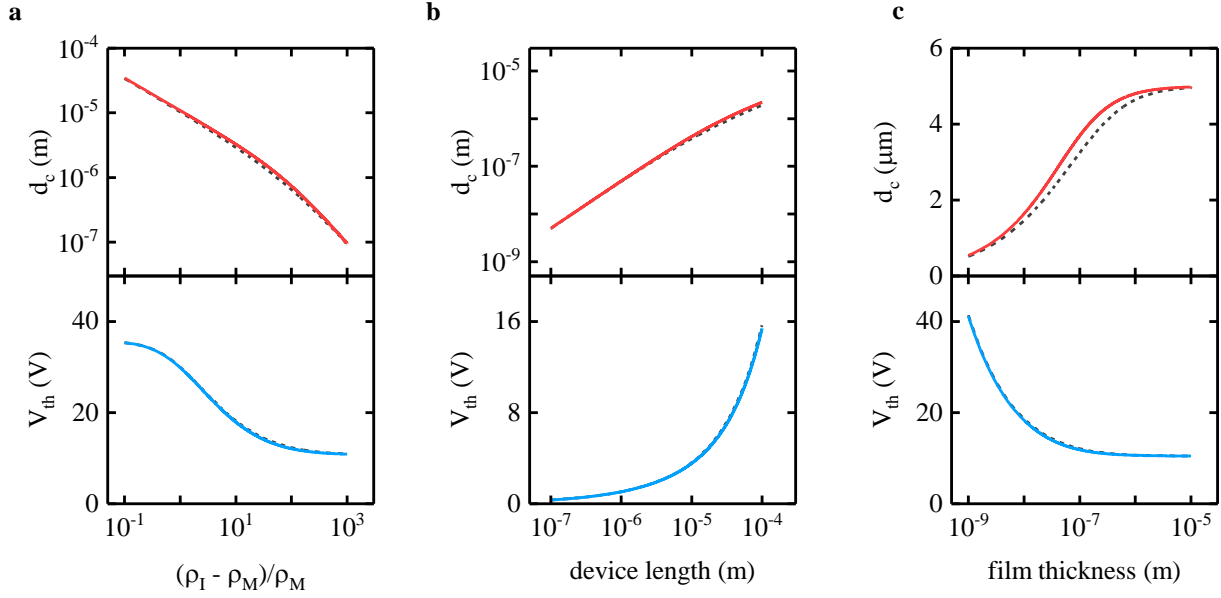

**Fig. S5.3.** Minimum insulating barrier size (top graphs) and threshold voltage (bottom graphs) dependence on resistivity ratio (**a**), device length (**b**) and film thickness (**c**). Continuous red and blue curves were obtained using equations (5.7-5.8). Dashed gray lines were obtained by solving numerically equation (5.3). Material and device parameters were set to the values presented in the text (p. 7), with the exception of the specific parameter for which $d_c$ and $V_{th}$ were calculated (i.e. resistivity ratio (**a**), device length (**b**) and film thickness (**c**)).

$$V_{th} \approx \sqrt{\frac{k_f L \rho_M (T_c - T_0)}{\alpha + 1} \left( \frac{k_s L}{k_f t} + 4 \left( \alpha + \sqrt{\alpha^2 + \alpha \frac{k_s L}{k_f t}} \right) \right)} \quad (5.8)$$

Fig. S5.3 shows the dependence of the minimum barrier size and threshold voltage, $d_c$ and $V_{th}$, on the resistivity ratio, device length, and film thickness. We plot the curves given by the approximate equations (5.7-5.8) (red and blue lines) and by numerically solving equation (5.3) using the $T_I \approx T_c$ and $T_M \approx T_0$ temperature estimations (grey dashed lines) and using the condition that the derivative of (5.3) is zero at $V_{th}$ (see the discussion on p.6). The material and device parameters were $T_c = 340$ K, $T_0 = 100$ K, $\rho_M = 2 \times 10^{-6}$ Ω·cm, $\alpha = 20$, $L = 100$ μm, $t = 20$ nm, $k_s = 5 \times 10^6$ W·K$^{-1}$·m$^{-2}$, $k_f = 3 \times 10^8$ W·K$^{-1}$·m$^{-2}$. These parameters give $d_c = 2.2$ μm and $V_{th} = 15.4$ V, which is very close to the experimentally observed values. The two approaches give almost the same results, which further supports the validity of the approximation introduced in (5.6). The analytical model predicts that the minimum insulating barrier size can be substantially reduced, potentially down to nanoscale, by selecting a material with large insulator/metal resistivity ratio or by reducing the device dimensions, length and film thickness. We note that because of the simplifications made in our model these results should be regarded as guidelines rather than exact predictions.

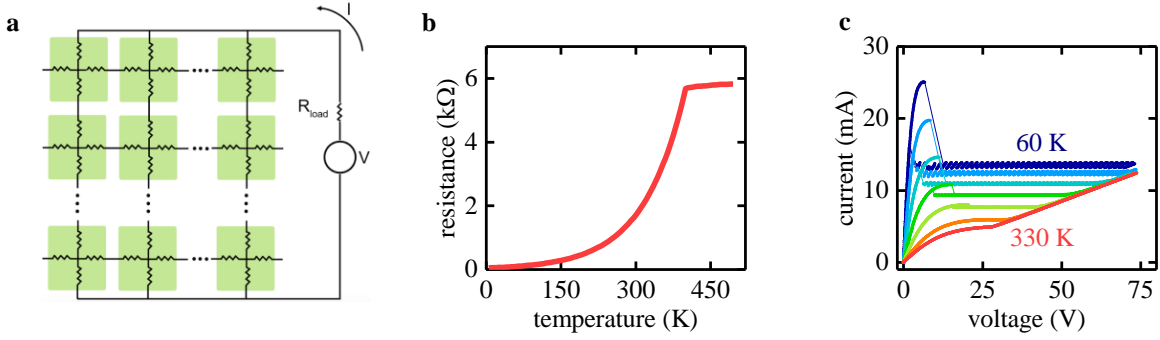

**Fig. S6. a,** A schematic of the resistor network used in simulations of the metal-to-insulator resistive switching. Resistor values at each node depend on local temperature given by equations (6.1-6.2). **b,** Simulated resistance-temperature dependence of the resistor network. **c,** Simulated voltage-controlled I-V curves in 60 – 330 K temperature range.

## Supporting Information 6

The computational analysis is based on a resistor network model as shown schematically in Fig. S6a. Each site of the 50×100 grid is represented by a 4-resistor node. The resistance of individual elements inside the nodes depends on temperature as

$$R(T) = \frac{R_0 e^\alpha}{m} + \frac{m-1}{m}\frac{R_0 e^\alpha}{1+e^{-\lambda(T-T_c)}} \tag{6.1}$$

where $R_0 = 18.5$, $m = 3$, $\lambda = 1.2 \times 10^{-4}$, $T_c = 340$, and

$$\alpha = \begin{cases} 0.012 \cdot T, & T < 400 \\ 0.012 \cdot 400, & T \geq 400 \end{cases} \tag{6.2}$$

These parameters were chosen to provide a semi-quantitative fit of the experimental $R(T)$ (see Fig. 1a in the main text). Fig. S6b shows the $R(T)$ plot of the full resistor network.

I-V curves and resistance maps were calculated in an iterative way. For a given applied voltage, the resistor network is solved to obtain local voltages $V_{ij}$ at each $(i, j)$ site. Then these voltages are used to update local temperatures $T_{ij}$ following the thermal diffusion equation:

$$\frac{dT_{ij}}{dt} = \frac{V_{ij}^2}{C_V R_{ij}} - \frac{k_h}{C_V}\left(5T_{ij} - T_{substrate} - \sum_{<kl>}^{1st\ neighboors} T_{kl}\right) \tag{6.3}$$

where $C_V = 2.0$ and $k_h = 0.4$. Using the local temperatures, local resistances are updated according to equations (6.1-6.2). The iterative process of solving the resistor network and updating the local temperatures and local resistances is repeated until a steady state is found.

Fig. S6c shows voltage-controlled I-V curves calculated in 60 – 330 K temperature range. Similar to experimental data (see main text Fig. 1c), we observe strongly nonlinear behavior and the appearance of an N-type NDR region at the temperatures below 300 K. To better illustrate how the nonlinearities in the I-V curves are related to the formation of an insulating barrier, we created several animations of the simulation results, which can be accessed at [LINK].